%% file: main.tex
\documentclass[pageno]{jpaper}

\usepackage[normalem]{ulem}
\usepackage{amsfonts}
\usepackage{amssymb}
\usepackage{amsmath}

\usepackage{booktabs} 
\usepackage{multirow}

\usepackage{algorithm}
\usepackage{algorithmic}

\newcommand{\algorithmiccontinue}{\textbf{continue}}
\newcommand{\CONTINUE}{\STATE \algorithmiccontinue}

\usepackage{caption} 
\newenvironment{sequation}{\vspace{-0.2cm}\begin{equation}\scriptsize}{\end{equation}\vspace{-0.2cm}}

\usepackage{hyperref}
\hypersetup{
    colorlinks = true,
}

\usepackage{enumitem}
\usepackage{comment}
\useunder{\uline}{\ul}{}

\usepackage{listings}
\usepackage{lstlinebgrd}
\usepackage{color}
\definecolor{dkgreen}{rgb}{0,0.6,0}
\definecolor{gray}{rgb}{0.5,0.5,0.5}
\definecolor{mauve}{rgb}{0.58,0,0.82}

\definecolor{bg}{rgb}{0.9,0.9,0.9}

\setlist[itemize,1]{itemsep=0.5pt,partopsep=0pt,parsep=\parskip, topsep=2pt, leftmargin=10pt,}
\setlist[enumerate,1]{itemsep=0.5pt,partopsep=0pt,parsep=\parskip, topsep=2pt, leftmargin=10pt,}

\newcommand{\framework}{AutoAccel}

\newcommand{\initslowdown}{314x}
\newcommand{\avgspeedup}{72x}
\newcommand{\avgimprove}{27,000x}

\makeatletter
\renewcommand*{\@seccntformat}[1]{\csname the#1\endcsname\hspace{0.5em}}
\makeatother

\begin{document}
\title{\framework: Automated Accelerator Generation and Optimization with Composable, Parallel and Pipeline Architecture}

\author{Jason Cong$^{1,2}$, Peng Wei$^{1}$, Cody Hao Yu$^{1,2}$, Peng Zhang$^{2}$ \\
 $^1$University of California, Los Angeles \\
 $^2$Falcon Computing Solutions, Inc.
}

\date{}
\maketitle
\thispagestyle{plain}

\input{0_abstract}
\input{1_intro}
\input{2_background}
\input{3_archi}
\input{4_model}

\input{5_framework}

\input{6_result}
\input{7_related}

\input{8_conclusion}

\bibliographystyle{plain}
\bibliography{reference} 

\end{document}

%% file: 0_abstract.tex
\begin{abstract}
CPU-FPGA heterogeneous architectures are attracting ever-increasing attention in an attempt to advance computational capabilities and energy efficiency in today's datacenters.
These architectures provide programmers with the ability to reprogram the FPGAs for flexible acceleration of many workloads.
Nonetheless, this advantage is often overshadowed by the poor programmability of FPGAs whose programming is conventionally a RTL design practice.
Although recent advances in high-level synthesis (HLS) significantly improve the FPGA programmability, it still leaves programmers facing the challenge of identifying the optimal design configuration in a tremendous design space.

This paper aims to address this challenge and pave the path from software programs towards high-quality FPGA accelerators.
Specifically, we first propose the composable, parallel and pipeline (CPP) microarchitecture as a template of accelerator designs. Such a well-defined template is able to support efficient accelerator designs for a broad class of computation kernels, and more importantly, drastically reduce the design space.
Also, we introduce an analytical model to capture the performance and resource trade-offs among different design configurations of the CPP microarchitecture, which lays the foundation for fast design space exploration.
On top of the CPP microarchitecture and its analytical model, we develop the {\framework} framework to make the entire accelerator generation automated.
{\framework} accepts a software program as an input and performs a series of code transformations based on the result of the analytical-model-based design space exploration to construct the desired CPP microarchitecture.
Our experiments show that the {\framework}-generated accelerators outperform their corresponding software implementations by an average of {\avgspeedup} for a broad class of computation kernels.

\end{abstract}

%% file: 1_intro.tex
\section{Introduction} \label{sec:intro}

Due to power and energy constraints, conventional general-purpose processors are no longer able to sustain the performance and energy improvement in commercial datacenters.
To overcome the inefficiency of homogeneous multicore systems, heterogeneous architectures that feature specialized hardware accelerators have been widely considered to be a promising paradigm.
In particular, field programmable gate arrays (FPGAs), which offer the potential of orders-of-magnitude performance/watt gains for a broad class of applications while retaining reconfigurability, attract increasing attention as a mainstream acceleration technology.
For example, both Microsoft and Baidu have incorporated FPGA-based accelerators in their datacenters to accelerate large-scale production workloads such as search engines~\cite{msft-catapult-isca14,msft-micro16} and neural networks~\cite{baidu-sda-hotchips14,msft-cnn-hotchips15}.
Amazon also introduced F1 instance~\cite{amazon-f1}, a compute instance equipped with FPGA boards, in its Elastic Compute Cloud (EC2).
Moreover, with the \$16.7 billion acquisition of Altera, Intel recently announced the Heterogeneous Architecture Research Platform (HARP)~\cite{harp}, which provides an FPGA and a Xeon processor in a single semiconductor package. Predictions have been made that as much as 30\% of datacenter servers will have FPGAs by 2020~\cite{harp-30}. 
This suggests that FPGAs could become a common component in future servers and could play an important role as primary computing resources~\cite{eriko2016fpl}. 

On the other hand, a major challenge in FPGA-based acceleration is programmability.
FPGA programming is generally recognized as an RTL (register-transfer level) design practice, which requires notable hardware expertise in designing accelerator microarchitectures such as controls, data paths, and finite state machines~\cite{rmm}. 
This makes the effort of FPGA programming prohibitive to most datacenter application developers.
It is even more challenging when the mainstream algorithm in an application domain is constantly evolving; i.e., an algorithm may have already been obsolete during the development process of its hardware accelerator.

Decades of research have focused on improving FPGA programmability. High-level synthesis (HLS)~\cite{cong11} that allows hardware designs to be described in high-level programming languages like C/C++ (such C/C++ programs for hardware designs are generally called hardware behavioral descriptions) is recognized as an encouraging approach.
In fact, a C program can even be compiled by state-of-the-art HLS tools like Xilinx SDAccel into a working FPGA circuit without any modification of the program itself.
However, a high-quality software program is generally far away from a high-quality hardware behavioral description due to the lack of proper consideration regarding the underlying FPGA architecture. Our experiments show that a software program, if naively treated as a hardware behavioral description, almost always leads to an FPGA accelerator that performs orders-of-magnitude worse than running the program on a modern CPU.
This is because HLS still leaves programmers to face the challenge of identifying the optimal design configuration among a tremendous number of choices, which in turn requires intimate knowledge of hardware intricacies to efficiently reduce the design space and obtain a high-quality solution in a reasonable time.
Consequently, to programmers HLS still presents a significant gap between a software program and a high-quality hardware behavioral description, which prevents the FPGA programmability from being further improved.

This paper presents a comprehensive approach to pave the path from a software program to a high-quality hardware behavioral description that 1) is functionally equivalent to the software program, and 2) leads to a high-performance FPGA accelerator.
The approach consists of three main stages.
The first stage, \textit{design space reduction}, aims to reduce the tremendous design space.
Specifically, we introduce the composable, parallel and pipeline (CPP) microarchitecture, a template of accelerator designs, as a specification of the program-to-behavioral-description transformation.
Such a carefully designed template fits for a variety of computation kernels and guarantees the quality of accelerator designs.
Also, with the CPP microarchitecture as the transformation specification, the design space is restricted to only configurations of that specific microarchitecture.
The second stage, \textit{automatic design space exploration}, realizes a near-optimal CPP microarchitecture configuration automatically with an analytical model and a machine-learning-based search engine.
With this near-optimal configuration, the third stage, \textit{automatic accelerator generation}, organizes a collection of code transformation primitives to transform the software program to the behavioral description of the desired CPP microarchitecture.
We develop the {\framework} framework to implement the proposed approach and make the entire accelerator generation process automated. In summary, this paper makes the following contributions:

\begin{itemize}
\item \textbf{The CPP microarchitecture.} By introducing this broadly applicable accelerator design template as the specification of program-to-behavioral-description transformation, we achieve the objective of drastically reducing the design space while preserving accelerator design quality.

\item \textbf{The analytical model.} This proposed model captures the performance and resource trade-offs among all design configurations of the CPP microarchitecture, laying the foundation for fast, automated design space exploration.

\item \textbf{The {\framework} framework.} {\framework} automates the entire accelerator generation process, provides datacenter application developers with a nearly push-button experience of FPGA programming, and thus substantially improves the FPGA programmability.

\item \textbf{Detailed evaluation.} We evaluate {\framework} via the MachSuite~\cite{machsuite} benchmark suite by proposing a metric to measure whether the qualities of {\framework}-generated accelerators reach optimality. 
We also evaluate the accuracy of the proposed analytical model using Xilinx SDAccel and the on-board execution.
\end{itemize}

Our experiments show that the {\framework}-generated accelerators outperform their corresponding software implementations by an average of {\avgspeedup} for the MachSuite computation kernels.


%% file: 2_background.tex
\section{Background} \label{sec:background}
A field-programmable gate array (FPGA) is an integrated circuit that contains an array of reprogrammable logic and memory blocks: lookup tables (LUTs), flip-flops (FFs), digital signal processing slices (DSPs) and block RAMs (BRAMs).
Connected through a hierarchy of reconfigurable interconnects, these blocks can be customized into different circuits to solve various computation problems.
Such hardware customizability allows FPGA circuits to avoid the significant overhead of the general-purpose microprocessors, resulting in orders-of-magnitude performance/watt gains for a broad class of workloads.

However, the FPGA programmability issue is a serious impediment against its adoption by datacenter application developers.
Section~\ref{subsec:hls} briefly describes state-of-the-art commercial HLS tools that represent the latest effort in improving the FPGA programmability through HLS.
The fact that such tools leave programmers to take full responsibility for performance optimization motivates our work.
In Section~\ref{subsec:merlin} we then introduce the Merlin compiler~\cite{merlin_product,merlin,merlin_islped}, a compilation framework that attempts to alleviate the burden of manual code optimization by providing a library of automated code transformation primitives.
While the Merlin compiler still relies on programmers to determine the optimal combination and parameters of the transformation operations, and thus does not substantially relieve the burden, its transformation library serves as a good preliminary tool for us to agilely implement automatic generation of the CPP microarchitecture.


\subsection{Commercial HLS Tools} \label{subsec:hls}

\begin{figure}[thb]
	\centering
	\includegraphics[scale=0.4]{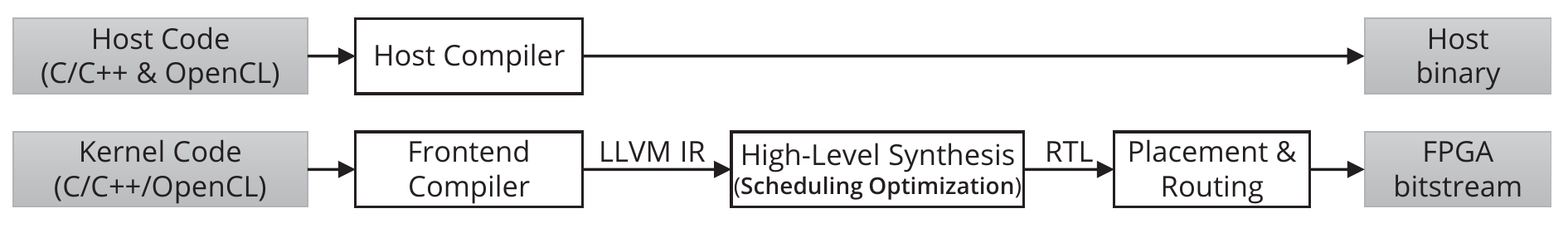} 
	\caption{Common Commercial HLS Tool Design Flow}
	\label{fig:commercial}
\end{figure}

Commercial HLS tools such as Xilinx SDAccel~\cite{sdaccel} and Intel FPGA SDK for OpenCL~\cite{intelsdk} have been widely used to fast prototype user-defined functionalities expressed in high-level languages (e.g., C/C++ and OpenCL) on FPGAs without involving register-transfer level (RTL) descriptions. 
The example design flow used by common commercial HLS tools is shown in Fig.~\ref{fig:commercial}.

Commercial HLS tools usually have a set of language extensions for users, such as C pragmas, that provide the guidances of memory organization and task scheduling to complement the missing information of static analysis while optimizing the design.
The language extensions are specified by the user at the source code level, but the core HLS code transformation and optimization happens at the intermediate representation (IR) level, indicating that the effectiveness of user guidances highly depends on its IR structure and front-end compiler.
It implies that two programs with the same functionality but different coding styles (leading to different IR structures) might result in a significant performance difference.
In fact, this difference can be up to several orders of magnitude based on our experiences.
As a consequence, programmers have to pay attention to every detail that may affect the generated IR structure, which often requires a profound understanding of the FPGA architecture and circuit design.



\subsection{Merlin Compiler} \label{subsec:merlin}

The Merlin compiler~\cite{merlin_product, merlin, merlin_islped} is a source-to-source transformation tool for FPGA acceleration based on the CMOST~\cite{cmost} compilation flow.
It provides a transformation library and a set of pragmas with prefix ``\texttt{\#pragma Accel}'' for developers to perform design optimization at the source-code level. Each pragma corresponds to a code transformation primitive, as listed in Table~\ref{tbl:merlin_pragmas}.

\input{tables/merlin_pragmas}

\begin{figure}[thb]
	\centering
	\includegraphics[scale=0.4]{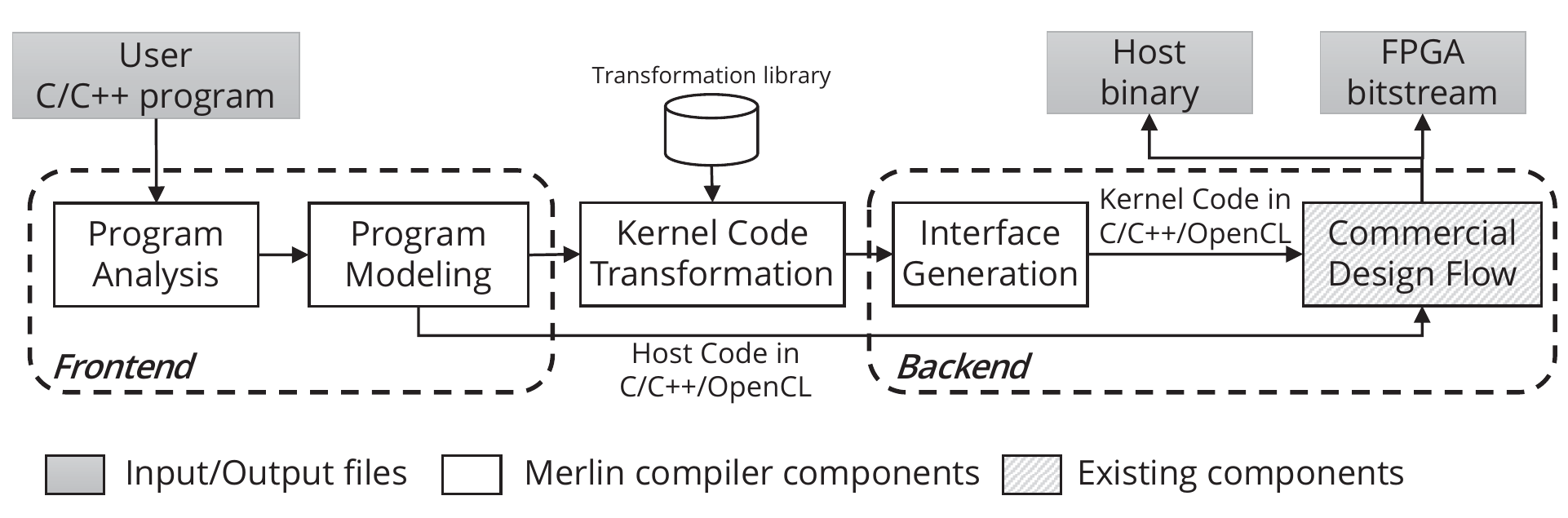} 
	\caption{The Merlin Compiler Execution Flow}
	\label{fig:merlin}
\end{figure}

Based on the transformation library, Fig.~\ref{fig:merlin} presents the Merlin compiler execution flow. It leverages the ROSE compiler infrastructure~\cite{rose} and polyhedral framework~\cite{polyframework} for abstract syntax tree (AST) analysis and transformation. The front-end stage analyzes the user program and separates host and computation kernel. The kernel code transformation stage then applies multiple code transformations according to user-specified pragmas. Note that the Merlin compiler will perform all necessary code reconstructions to make a transformation effective. For example, when performing loop unrolling, the Merlin compiler not only unrolls a loop but also conducts memory partitioning for the sake of avoiding bank conflict~\cite{cong-todaes11}. Finally, the back-end stage takes the transformed kernel and uses the HLS tool to generate the FPGA bitstream.


Compared to the pure HLS solution, the Merlin compiler further improves the FPGA programmability by making design optimization ``semiautomatic'': instead of manually reconstructing the code to make one optimization operation effective, programmers now can simply place a pragma and let the Merlin compiler do the necessary changes.
However, programmers still have to identify the best combination and parameters among these operations, i.e., manually searching in an exponential design space.


%% file: tables/merlin_pragmas.tex
\begin{table}[thb]
\scriptsize
\centering
\caption{Merlin Compiler Code Transformations}
\label{tbl:merlin_pragmas}
\begin{tabular}{|l|l|l|l|}
\hline
Transformation & Target & Parameters & Description \\ \hline\hline
\multirow{2}{*}{Data tiling} & Loop & tilesize=$S$ & \begin{tabular}[c]{@{}l@{}}Tile the loop and create on-chip \\ buffers to cache the data with size $S$.\end{tabular} \\ \cline{2-4} 
 & \multicolumn{3}{l|}{Example: \#pragma Accel data\_tiling tilesize=16} \\ \hline
\multirow{2}{*}{\begin{tabular}[c]{@{}l@{}}Memory \\ Coalescing \end{tabular}} & Buffer & bitwidth=$b$ & Pack DRAM buffer to $b$ bits. \\ \cline{2-4} 
 & \multicolumn{3}{l|}{Example: \#pragma Accel bitwidth variable=buf factor=512} \\ \hline
\multirow{2}{*}{Pipeline} & Loop & N/A & \begin{tabular}[c]{@{}l@{}}Create a coarse- or fine-grained \\ pipeline (dataflow).\end{tabular} \\ \cline{2-4} 
 & \multicolumn{3}{l|}{Example: \#pragma Accel pipeline} \\ \hline
\multirow{2}{*}{Parallelism} & Loop & factor=$N$ & \begin{tabular}[c]{@{}l@{}}Tile the loop and create $N$ \\ processing elements (PEs).\end{tabular} \\ \cline{2-4} 
 & \multicolumn{3}{l|}{Example: \#pragma Accel parallel factor=4} \\ \hline
\end{tabular}
\end{table}

%% file: 3_archi.tex
\section{Accelerator Design Template} \label{sec:archi}

This section presents the details of the \textit{design space reduction} stage of the proposed approach.
In general, our solution is to introduce an accelerator design template as the specification of the transformation from software programs to hardware behavioral descriptions.
A software program will only be transformed to a hardware behavioral description of this introduced template, so the design space is restricted to only configurations of the template.
As a result, the design space is drastically reduced (see Section~\ref{sec:model} for design space definition).
Meanwhile, this template ought to be applicable for a variety of computation kernels, and guarantees the accelerator design quality once a kernel fits into the template.
Section \ref{subsec:analysis} and \ref{subsec:arch} present our proposed accelerator design template, the composable, parallel and pipeline (CPP) microarchitecture, as well as showing how the CPP microarchitecture is derived.
Section \ref{subsec:ds} discusses the applicability of the CPP microarchitecture for various computation kernels.

\subsection{Obstacles Towards Efficient Behavioral Description} \label{subsec:analysis}

We derive the CPP microarchitecture by conducting an analysis on the major obstacles from a software program towards an efficient hardware behavioral description.
Specifically, we start from a collection of computation kernels, straightforwardly treat their software implementations\footnote{The computation kernels and their software implementations are from the MachSuite benchmark suite~\cite{machsuite} (see Section~\ref{subsec:exp_setup}).} as behavioral descriptions, feed such naive behavioral descriptions into Xilinx SDAccel, and identify the microarchitectural inefficiencies of the generated FPGA accelerators.
Such inefficiencies represent the obstacles towards efficient behavioral descriptions.

We use the NW (Needleman-Wunsch algorithm) benchmark (see Section \ref{subsec:exp_setup}) as an example for demonstration and discussion.
The NW benchmark processes a series of genome sequence alignment jobs, each with a pair of 128-entry sequences as input and a pair of 256-entry sequences as output.
The alignment engine applies the Needleman-Wunsch algorithm, a dynamic programming algorithm with quadratic time complexity, to the input sequences, and generates the optimal post-aligned sequences given a predefined scoring system~\cite{nwalgorithm}.

\begin{figure}[t]
\begin{minipage}{4.5cm}
\begin{lstlisting}[basicstyle={\scriptsize\ttfamily}]
void engine(...) {
  int M[129][129];
  ...
loop1: for(i=0; i<129; i++) {M[0][i]=...}
loop2: for(j=0; j<129; j++) {M[j][0]=...}
loop3: for(i=1; i<129; i++) {
    for(j=1; j<129; j++) {...
      M[i][j]=...
  }}
  ...
}
void kernel(char seqAs[], char seqBs[],
       char alignedAs[], char alignedBs[]) {
  for (int i=0; i<NUM_PAIRS; i++) {
    engine(seqAs+i*128, seqBs+i*128,
       alignedAs+i*256, alignedBs+i*256);
}}
\end{lstlisting}    	
\end{minipage}
\begin{minipage}{3.5cm}
\raggedright
\includegraphics[scale=0.35]{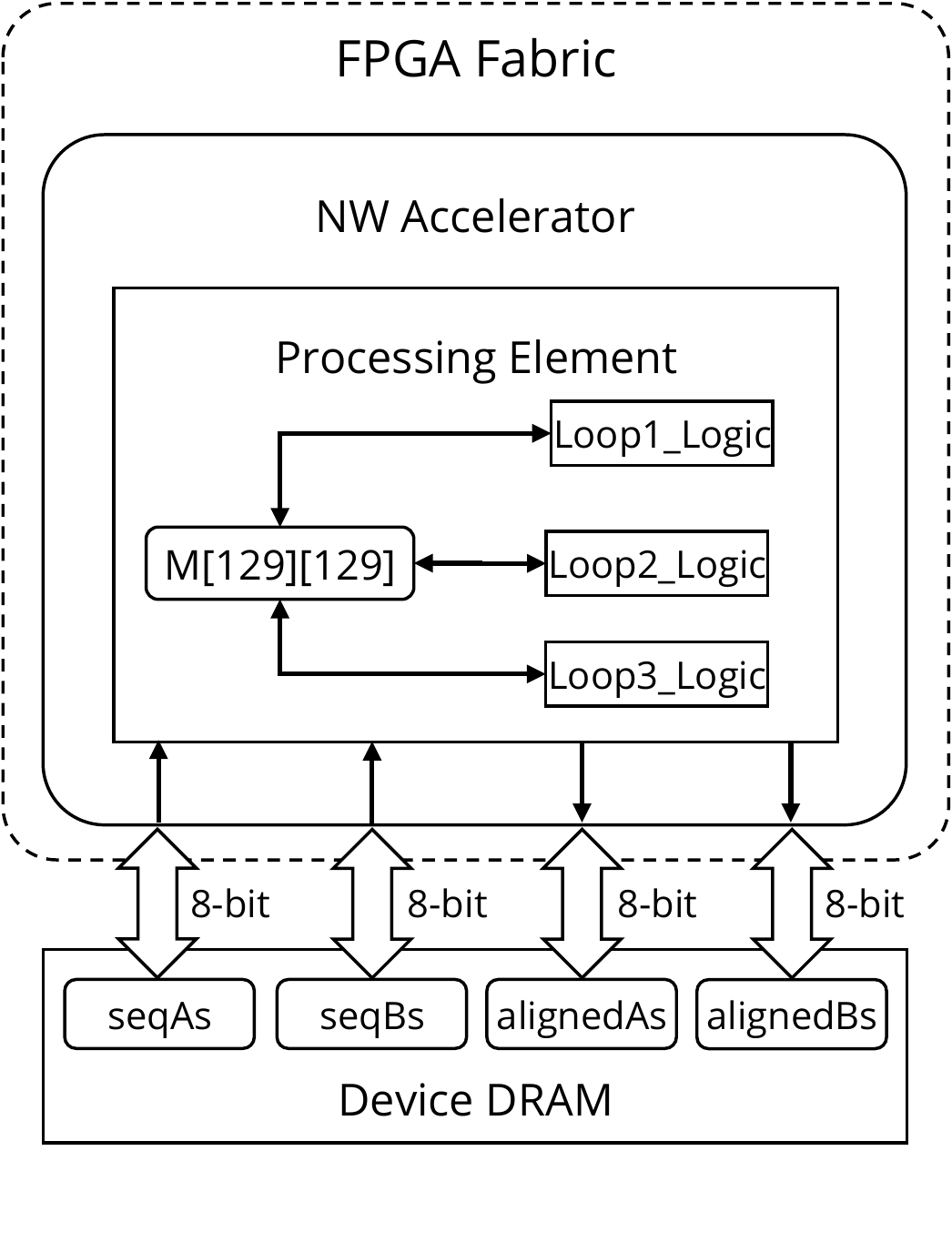}
\end{minipage}
\caption{NW Kernel and the Corresponding Architecture}
\label{fig:baseline}
\end{figure}

Fig.~\ref{fig:baseline} presents the NW code snippet and the microarchitecture of the FPGA accelerator generated by naively feeding the NW code into Xilinx SDAccel.
Our experiments show that this accelerator performs 92x slower than a single CPU core.
We dig into the implementation inefficiencies of the NW benchmark that cause such poor performance as follows.

\noindent\textbf{Inefficiency \#1: Inefficient off-chip transaction.} The \texttt{kernel} function is the top-level function of the NW benchmark and defines the entire accelerator.
Its arguments---\texttt{seqAs}, \texttt{seqBs}, \texttt{alignedAs} and \texttt{alignedBs} that correspond to the original sequence pairs and the aligned sequence pairs---define the input and output buffers that reside in the off-chip DRAM of the FPGA board. The FPGA accelerator connects to these off-chip buffers through AXI channels.
The data width of each AXI channel is eight bits, inferred from the data type of the corresponding argument (8-bit \texttt{char} type in the NW case).
As a result, the off-chip data transaction throughput is only one byte/cycle for each channel, or four byte/cycle aggregately, while state-of-the-art CPU-FPGA platforms typically support 64 byte/cycle off-chip communication throughput.

\noindent\textbf{Inefficiency \#2: No data caching.} No data caching module is presented in the microarchitecture, with the result that every data access goes through the off-chip DRAM.

\noindent\textbf{Inefficiency \#3: Sequential loop scheduling.} The \texttt{kernel} function body is a loop statement that iteratively traverses every sequence pair through the \texttt{engine} function that defines the hardware \texttt{engine} module. 
In the presented microarchitecture, the \texttt{engine} module accepts and processes only one sequence pair at a time, despite the fact that these sequence pairs are independent of each other and thus can be processed in parallel or pipeline.
Worse still, all loops presented in the NW kernel are scheduled to be processed sequentially, regardless of whether one is able to be mapped to a parallel or pipeline circuit.\footnote{The latest Xilinx flow starts to perform loop pipelining automatically, but only for simple loop statements.}

\noindent\textbf{Inefficiency \#4: Inefficient on-chip memory utilization.} The major computation of the NW algorithm is to generate a two-dimension score matrix.
The \texttt{engine} function therefore includes a local two-dimensional array, \texttt{M}, to store the matrix, and some loop statements to calculate the values of the matrix elements.
In the presented microarchitecture, the array \texttt{M} is mapped to an on-chip BRAM buffer that has only one write port, implying that even if the algorithm has the potential to generate multiple matrix element values per cycle, the BRAM buffer is not able to fulfill this potential because only one value can be written into the buffer in each cycle.

These inefficiencies, though demonstrated only in the NW example, are present in all MachSuite benchmarks and represent the major obstacles from software programs to high-quality hardware behavioral descriptions.
The CPP microarchitecture is thus derived to resolve these inefficiencies.

\subsection{CPP Microarchitecture} \label{subsec:arch}
\begin{figure}[thb]
	\centering
	\includegraphics[scale=0.30]{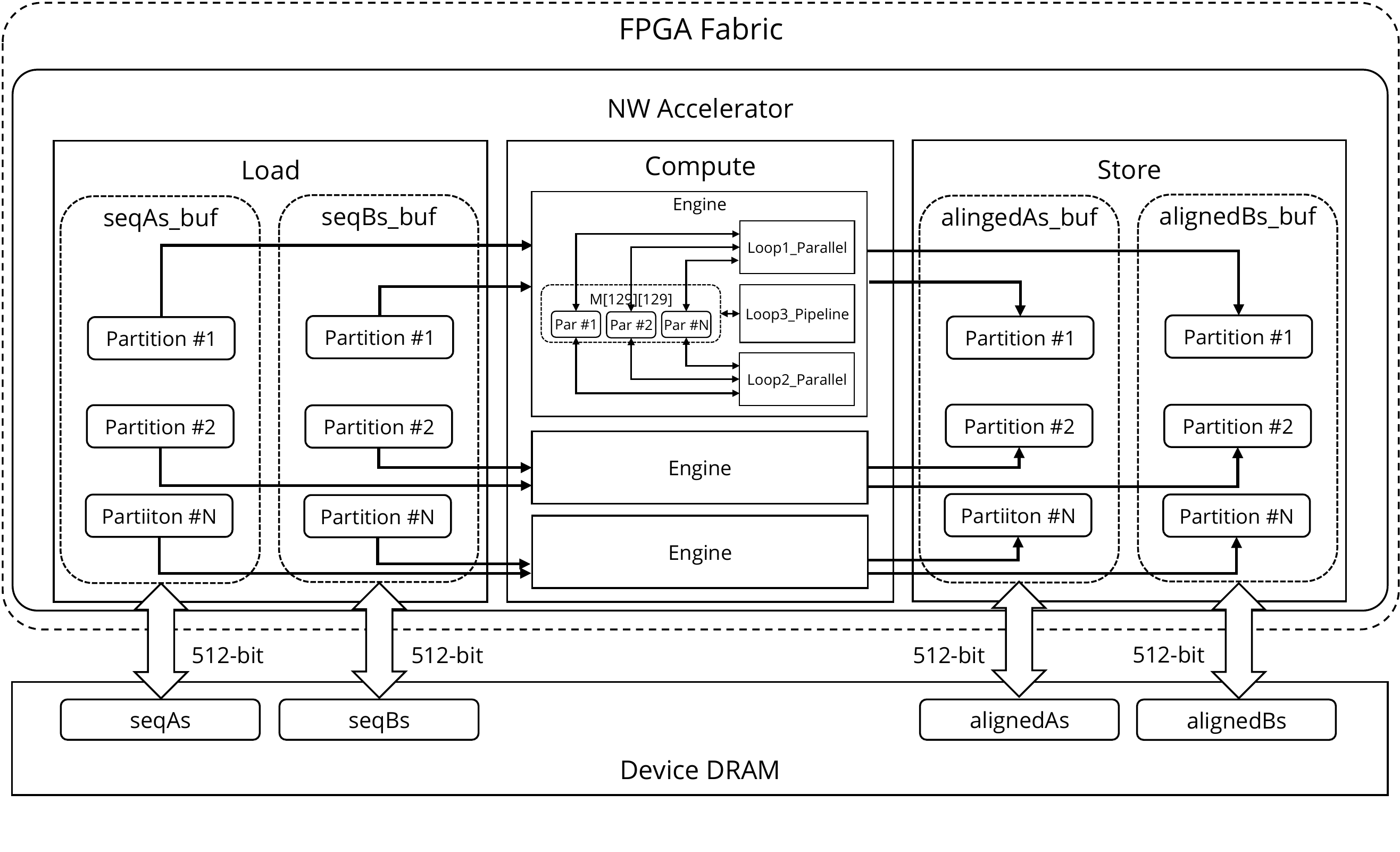} 
	\caption{NW Accelerator under CPP Microarchitecture}
	\label{fig:overall}
\end{figure}

The composable, parallel and pipeline (CPP) microarchitecture is proposed as a template of accelerator designs and a specification of the program-to-behavioral-description transformation.
It includes a series of features to address the inefficiencies in the previous section.
In the following text we continue to use the NW benchmark as an example to demonstrate the CPP microarchitecture along with its key features, as shown in Fig.~\ref{fig:overall}.

\noindent\textbf{Feature \#1: Coarse-grained pipeline with data caching.} Fig.~\ref{fig:overall} illustrates the NW accelerator design under the CPP microarchitecture.
The overall CPP microarchitecture is a coarse-grained pipeline that consists of three stages: \texttt{load}, \texttt{compute} and \texttt{store}.
The \texttt{kernel} function in the NW source code only corresponds to the \texttt{compute} module instead of defining the entire accelerator.
The input sequence pairs are processed tile by tile, i.e., iteratively loading a certain number of sequence pairs into on-chip buffers (Stage \texttt{load}), aligning these pairs (Stage \texttt{compute}), and storing the post-aligned pairs back to DRAM (Stage \texttt{store}).
Different tiles are processed in pipeline since they are independent from each other.
This feature addresses \textit{inefficiency \#2} because off-chip data movement only happens in the \texttt{load} and \texttt{store} stages, leaving the data accesses of computation completely on chip.

The \texttt{load} and \texttt{store} modules connect to two input and output DRAM buffers, respectively, through AXI channels. The data widths of the AXI channels are decoupled from the type sizes of the top-level function arguments. Hence, the off-chip bandwidth can potentially reach the highest physical bandwidth of the CPU-FPGA platform. Also, the \texttt{load}-\texttt{compute}-\texttt{store} pipeline improves the effective bandwidth of the accelerator by overlapping communication with computation. Consequently, \textit{inefficiency \#1} is addressed as well.

\noindent\textbf{Feature \#2: Loop scheduling.} The CPP microarchitecture tries to map every loop statement presented in the computation kernel function to either 1) a circuit that processes different loop iterations in parallel, 2) a pipeline where the loop body corresponds to the pipeline stages, or 3) a combination of both.
As for the NW example, the loop statement in the \texttt{kernel} function is mapped to a set of \texttt{engine} modules to process the sequence pairs in parallel.
Moreover, the loop statements in the \texttt{engine} function are mapped to parallel and pipeline circuits as well.
This resolves \textit{inefficiency \#3}.


\noindent\textbf{Feature \#3: On-chip buffer reorganization.} 
In the CPP microarchitecture, all the on-chip BRAM buffers are partitioned to meet the port requirement of parallel circuits, where the number of partitions of each buffer is determined by the duplication factor of the parallel circuit that connects to the buffer. This feature is used for resolving \textit{inefficiency \#4}. In the NW example, the on-chip buffers that cache the input and output sequence pairs are partitioned into multiple segments, each segment feeding one \texttt{engine} module.
The local buffer \texttt{M} that stores the score matrix is also partitioned to allow parallel read and write transactions.

In summary, the CPP microarchitecture guarantees the quality of accelerator designs by providing corresponding features to address the inefficiencies.
However, it is not applicable to all kinds of computation kernels with various data processing patterns. The following section discusses the applicability of the CPP microarchitecture for various computation kernels.

\subsection{Applicability for Computation Kernels} \label{subsec:ds}

The CPP microarchitecture features a load-compute-store coarse-grained pipeline, which requires the computation kernel to process input data block by block.
Meanwhile, the size of each block is required to be less than a few megabytes in order to be entirely cached on chip.
As a consequence, the CPP microarchitecture favors the computation kernels with regular data-level parallelism, like streaming or batch processing programs with the MapReduce~\cite{mapreduce} pattern.
On the contrary, it does not fit well for the computation kernels featuring extensive random accesses on a large memory footprint, such as PageRank~\cite{pagerank} and and the breadth-first search (BFS) algorithm.

%% file: 4_model.tex
\section{Analytical Model} 
\label{sec:model}

Another advantage of using CPP microarchitecture is to have a clear design space.
This section presents our CPP microarchitecture analytical model that estimates the execution cycles and resource consumptions of these configurations; this lays the foundation for the \textit{automatic design space exploration} stage of the proposed approach.

Unlike most existing models~\cite{gao-fpga16,dhdl,raghu-asplos,fpga-opencl,zhongdesign} that analyze the source program directly, many parameters of our proposed model are obtained from the HLS synthesis reports of a few design points.
This feature enables our model to capture most scheduling optimizations performed by the HLS tool.
As we will show in Section \ref{sec:exp}, the proposed model has less than a $5\%$ error rate compared to the HLS report.

\subsection{Performance Modeling}
\label{sec:model_perf}
The performance model estimates an accelerator's overall execution cycle ($C$) through Eq.~\ref{eq:perf}:

\begin{sequation}
\label{eq:perf}
C = max(C_{l} + C_{s}, C_{c})
\end{sequation}

\noindent where $C_{l}$, $C_{c}$ and $C_{s}$ denote the cycles of the load, compute and store modules, respectively. Since the load and store modules share the off-chip bandwidth and are together overlapped with the compute module in our experimental platform, we make a maximum operation between the cycles of the load/store modules and that of the compute module.

The execution cycles of the load, compute and store modules, as well as all of their submodules, can be quantified as the total cycles of all its loops ($C_{loop}$), submodules ($C_{mod}$) and standalone logic ($C_{r}$), as shown Eq.~\ref{eq:peperf}.

\begin{sequation} \label{eq:peperf}
C_{mod}(M) = \sum_{i \in M.loops}{C_{loop}(i)} + \sum_{m \in M.mods}{C_{mod}(m)} + C_{r}(M)
\end{sequation}

\noindent where $C_{r}$ is obtained from the HLS report. 

Then we model the loop execution. Although a loop statement can be scheduled in pipeline, parallel or the combination of both, the first two schedules can be treated as special cases of the last one, and can together be modeled as Eq.~\ref{eq:loopperf}:

\begin{sequation} \label{eq:loopperf}
C_{loop}(L) = C_{iter}(L) + II(L) \times \frac{TC(L)}{UF(L)}
\end{sequation}

\noindent where $C_{iter}$, $II$, $TC$ and $UF$ denote the iteration latency, initiation interval, trip count and unroll factor, respectively. 
$II$ and $TC$ are obtained from the HLS report; $UF$ is a design parameter that needs to be explored.

Subsequently, we break down and model the loop iteration in Eq.~\ref{eq:loopiter}, where the loop iteration latency is composed of the total cycles of all their sub-loops, submodules and standalone logic.

\begin{sequation} \label{eq:loopiter}
C_{iter}(L) = \sum_{i \in L.loops}{C_{loop}(i)} + \sum_{m \in L.mods}{C_{mod}(m)} + C_{r}(L)
\end{sequation}

Eq.~\ref{eq:peperf} and Eq.~\ref{eq:loopiter} reflect the architecture hierarchy with nested modules and loops.
The proposed model recursively traverses all the loops and modules until a loop or module does not contain any sub-structures.
In addition, we can find that Eq.~\ref{eq:peperf} and Eq.~\ref{eq:loopiter} are almost identical. This is because the loop iteration can be treated as a special ``module'' and modeled in the same way for both performance and resource.
Hence, we omit the loop iteration breakdowns in the following resource models.


\subsection{Resource Modeling} 
\label{sec:model_res}
The resource models estimate the consumptions of the four FPGA on-chip resources: BRAMs, LUTs, DSPs and FFs.
As the DSP model is relatively straightforward and the FF model is similar to the LUT model, we only demonstrate the BRAM and LUT models in this section.

\noindent \textit{\textbf{BRAM modeling}}: The BRAM consumption of a hardware module consists of the BRAM blocks used by all its local buffers ($R^{mem}_{buf}$) and those used by all its submodules ($R^{mem}_{mod}$), as shown in Eq.~\ref{eq:bram}:

\begin{sequation} \label{eq:bram}
R^{mem}_{mod}(M) = \sum_{b \in M}{R^{mem}_{buf}(b)} + \sum_{m \in M.mods}{R^{mem}_{mod}(m) \times DF(m)}
\end{sequation}

\noindent where $DF(m)$ is the duplication factor of submodule $m$ which is equivalent to the unroll factor of the loop that includes this submodule. We use ``duplication factor'' instead of ``unroll factor'' since the former is a better fit for depicting hardware modules and the latter is more suitable for describing loop statements.

Then we model the BRAM consumption of on-chip buffers.
A buffer's BRAM consumption is determined by three factors: 1) partition factors on all dimensions, $\prod_{d \in dim(B)}{PF(d)}$, 2) the size of each partition, $\lceil\frac{S(B)}{\prod_{d}{PF(d)}}\rceil$, and 3) the bit-width of the buffer, $bw(B)$, as shown in Eq.~\ref{eq:bufbram}:

\begin{sequation} \label{eq:bufbram}
R^{mem}_{buf}(B) = \prod_{d \in dim(B)}{PF(d)} \times V\left(\lceil\frac{S(B)}{\prod_{d}{PF(d)}}\rceil, bw(B)\right)
\end{sequation}

Eq.~\ref{eq:bufbram} adopts a function $V(s, b)$ to calculate the BRAM consumption of a single partition. The two parameters are the size and the bit-width of the partition. Eq.~\ref{eq:singlebufbram} presents its expression:

\begin{sequation} \label{eq:singlebufbram}
V(s, b) = \lceil\frac{s}{N_{blk}(b) \times S_{unit}}\rceil \times N_{blk}(b)
\end{sequation}


\noindent where $S_{unit}$ denotes the size of a BRAM block that is a platform-dependent constant.
$N_{blk}(b)$ is a function that calculates the minimum number of BRAM blocks needed to compose a BRAM buffer with bit-width $b$. Eq.~\ref{eq:blkbram} shows its expression, where $b_{phy}$ is a platform-dependent constant that represents the largest supported bit-width of a BRAM building block.

\begin{sequation} \label{eq:blkbram}
N_{blk}(b) = \lceil\frac{b}{b_{phy}}\rceil
\end{sequation}

\noindent \textit{\textbf{LUT modeling}}: The LUT consumption of a hardware module ($R^{lut}_{mod}$) is composed of the number of LUTs used by all loops, submodules, BRAM buffers (for control logic) and the standalone logic: 

\begin{sequation} \label{eq:lut}
\begin{aligned}
R^{lut}_{mod}(M) &= \sum_{l \in M.loops}{R^{lut}_{iter}(l) \times UF(l)} + \sum_{b \in M.bufs}{R^{lut}_{buf}(b)} \\
&+ \sum_{m \in M.mods}{R^{lut}_{mod}(m) \times DF(m)} + R^{lut}_{r}(M)
\end{aligned}
\end{sequation}

\noindent where $R^{lut}_{iter}$ depicts the LUT consumption of the loop iteration that is, again, treated and modeled as a special ``module.''
$R^{lut}_{r}$ denotes the LUT consumption of the standalone logic and is obtained from \textit{two} HLS reports.

We then model the LUT consumption of on-chip buffers ($R^{lut}_{buf}$).
It can be decoupled into two parts: 1) the control ($R^{lut}_{ctrl}$) and data ($R^{lut}_{data}$) signals of each BRAM partition, and 2) the $k$-to-1 multiplexer ($R^{lut}_{mux}(k)$) that selects the desired data from all the partitions, as shown in Eq.~\ref{eq:buflut}:  

\begin{sequation} \label{eq:buflut}
R^{lut}_{buf}(B) = R^{mem}_{buf}(B) \times (R^{lut}_{ctrl} + R^{lut}_{data}) + R^{lut}_{mux}\left(\prod_{d \in dim(B)}{PF(d)}\right) \times bw(B)
\end{sequation}
\begin{sequation} \label{eq:mux}
R^{lut}_{mux}(k) = \sum_{i=1}^{\lceil log_4{k} \rceil}{\lceil \frac{k}{4^i} \rceil}
\end{sequation}

\noindent 
where $R^{lut}_{ctrl}$ and $R^{lut}_{data}$ are obtained from the HLS report, and $R^{lut}_{mux}$ can be calculated via Eq.~\ref{eq:mux}.
We can also see that the LUT consumption of a buffer depends on its BRAM usage.

Based on the proposed model, the design space of the CPP microarchitecture is composed of 1) the capacity and bit-width of every on-chip buffer, and 2) the unroll factor of every loop, as indicated in Table~\ref{tbl:ds}.
Unfortunately, the proposed model is neither linear nor convex, and therefore not able to be mathematically solved in polynomial time.
Hence, we implement automatic design space exploration by leveraging a machine-learning-based search engine that is able to greatly reduce the number of search iterations needed to reach a near-optimal solution. This, together with the {\framework} framework, will be presented in the following section.

\input{./tables/design_space}

%% file: tables/design_space.tex
\begin{table}[thb]
\centering
\caption{The CPP Microarchitecture Design Space}
\label{tbl:ds}
{\scriptsize
\begin{tabular}{|l|l|}
\hline
Name & Design Space \\ \hline\hline
Buffer size & $\left\{s \mid s = S(B) \in \mathbb{B}, 0 < s < 4M\right\}$ \\ \hline
Buffer bit-width & $\left\{b \mid b = bw(B) \in \mathbb{B}, 8 < b = 2^{n} < 512\right\}$ \\ \hline
Loop unroll factor & $\left\{u \mid u = UF(u) \in \mathbb{L}, 1 < u < TC(L)\right\}$ \\ \hline
\end{tabular}
}
\end{table}

%% file: 5_framework.tex
\section{{\framework} Framework} \label{sec:framework}
In this section we present the {\framework} framework that takes a nested loop\footnote{Computation kernels with multiple nested loops can be decoupled into multiple sub-kernels, each corresponding to a CPP microarchitecture. Existing work~\cite{huang2014} has extensively studied how to connect multiple accelerators through FIFO channels with efficient inter-accelerator communication.} in C as input and performs a series of transformations to produce a high-quality FPGA accelerator with the CPP microarchitecture.
{\framework} is built on top of the Merlin compiler and uses its transformation library to construct the CPP microarchitecture.

\begin{figure}[thb]
	\centering
	\includegraphics[scale=0.42]{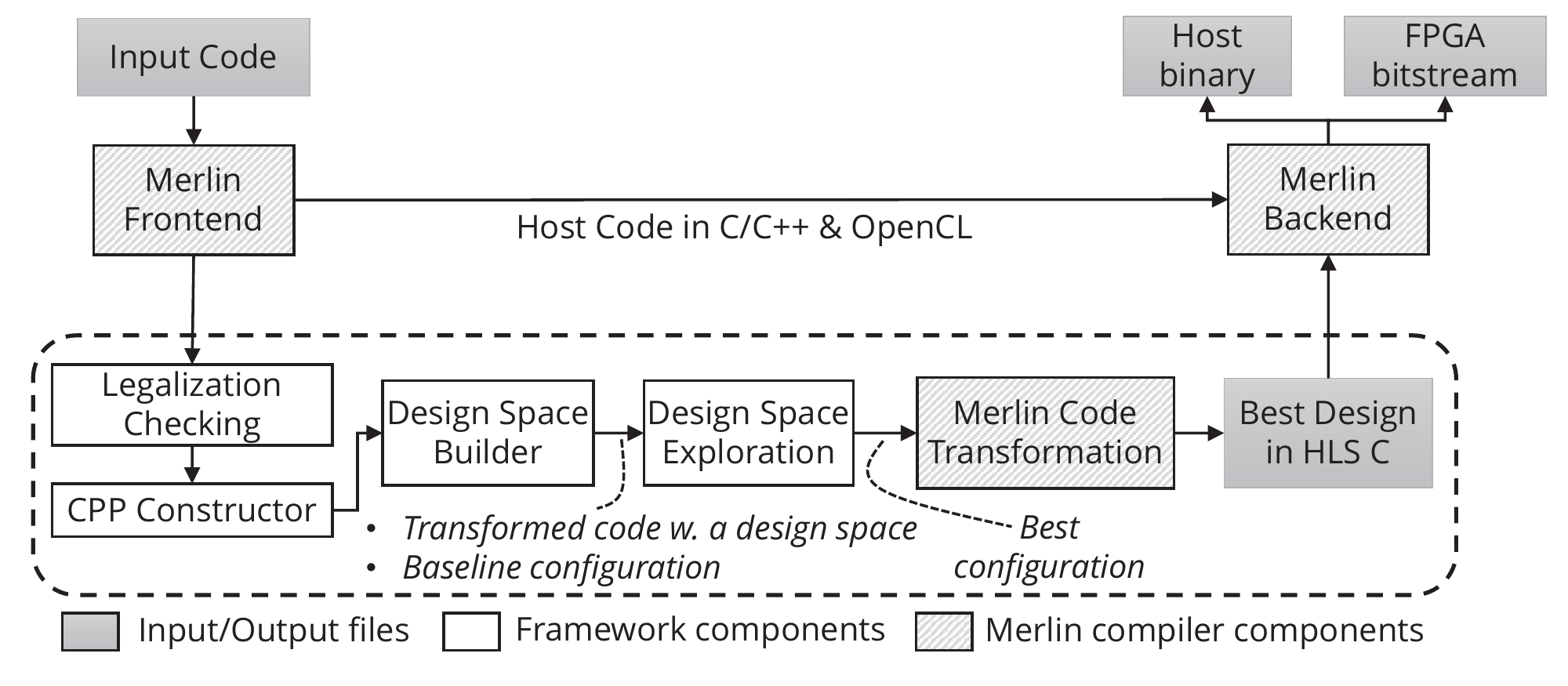} 
	\caption{{\framework} Framework Overview}
	\label{fig:framework}
\end{figure}

Fig.~\ref{fig:framework} illustrates the overall flow of the {\framework} framework. The input program is first evaluated by the legalization checking to determine whether it fits into the CPP microarchitecture. Next, we implement a CPP microarchitecture constructor to refactor the input program to a hardware behavioral description of the CPP microarchitecture. Subsequently, a design space builder is developed to identify the design space via static code analysis. After the design space has been built, we introduce a design space explorer with our proposed analytical model to realize the best design specification in minutes. Finally, we refactor the behavioral description code again by applying the best design specification to generate the desired accelerator design. This design can be directly fed into Xilinx SDAccel to derive a high-quality accelerator bitstream. In the remainder of this section we present the detailed implementation of each component.

\subsection{Legalization Checking}
Since {\framework} does not require any user modification of the input computation kernel code, the goal of legalization checking is to evaluate whether the input kernel is able to be mapped to the CPP microarchitecture. We briefly describe the evaluation points of the {\framework} built-in legalization checking algorithm as follows:

\textbf{\textit{Kernel size}}. The resource requirements of generated designs cannot exceed the capacity of single FPGA fabric. This can be evaluated by running HLS with the basic configuration. 

\textbf{\textit{Task-dependent data chunk length}}. A task-dependent array is an array that is traversed by the PE-loop so that every PE will use a different chunk of data.
To achieve the most efficient parallelism and pipeline scheduling in the CPP microarchitecture, the on-chip scratchpad memory is partitioned for every PE to avoid writing conflicts.
For example, the string length of NW kernel in Fig.~\ref{fig:baseline} is always 128, so it can be processed by {\framework}.
However, if the size of the task data chunk is determined dynamically, {\framework} cannot statically allocate a certain memory size to each PE; this results in the failure of legalization checking.

\textbf{\textit{Task-independent data size}}.
A task-independent array, on the other hand, is an array that is accessed by all PEs.
For example, in the breadth-first search (BFS) implementation of the MachSuite benchmark~\cite{machsuite}, the array that stores the tree is task-independent, because every PE might access any part in the array so that it cannot be partitioned regularly. As a result, it is better to duplicate task-independent arrays in on-chip memory for each PE to guarantee the efficiency. In case the array is too large to be stored in on-chip memory, the kernel fails to pass the legalization checking.

We perform legalization checking by traversing an abstract syntax tree (AST). We analyze the iteration domain to reason kernel accessed data size by the polyhedral analysis from \cite{Pouchet-fpga13}.

\subsection{CPP Microarchitecture Construction and Design Space Establishment} \label{subsec:sda_create}
{\framework} makes use of the transformation library of the Merlin compiler to preprocess user input code to fit the CPP microarchitecture. To constrain a design space when constructing the CPP microarchitecture, we use static analysis and a polyhedral model to collect the necessary information (e.g., loop trip count, maximal buffer size, bit-width, etc). Instead of specifying an integer number in Merlin pragmas for a certain configuration (e.g., data tiling size), we define an expression ``\texttt{auto(min, max, inc)}'' to represent a set of design points. In the expression, \texttt{min} and \texttt{max} indicate the range while \texttt{inc} specifies the incremental operator from the minimum value to the maximum value. 
We currently support two incremental operators: 1) \texttt{seq} that represents the ``$+1$'' increment, and 2) \texttt{pow2} that represents the ``$\times 2$'' increment.
This expression will be replaced with a specific integer of the best configuration after the design space exploration (DSE).

We now introduce the transformation operations used to construct the CPP microarchitecture. Again, the NW benchmark is used as an example to demonstrate the transformation flow, as shown in Code~\ref{list:final}.
The first three transformations are data tiling, coarse-grained pipeline and processing element duplication.

\noindent\textbf{1. Data tiling: } The transformation first tiles a sub-loop in the nested loop and creates a set of on-chip buffers for data caching. Then it instruments the code for establishing efficient off-chip data communication by enabling memory burst. The transformed code corresponds to lines 33-51 in Code~\ref{list:final}.


Since the CPP microarchitecture decouples the off-chip memory communication from computation, the analytical model does not cover the design points with different data tiling granularity. To solve this problem, we find the best design point of all possible data tiling granularities that are reported by the legalization checking algorithm in parallel. We plan to include data tiling granularity into the design space in the future.

\noindent\textbf{2. Coarse-grained pipeline: } After the data tiling, we apply the coarse-grained pipeline transformation that encapsulates load, compute, store into three functions to draw the boundaries between pipeline stages (lines 41-51). Subsequently, the transformation duplicates on-chip buffers created by step 1 and interleaves all of them by enabling double buffering.

\noindent\textbf{3. Processing element duplication: } The next step is to enable parallel computing. We apply the parallelism transformation to the compute stage in the tiled nested loop (lines 20-24). This creates multiple homogeneous processing elements (PEs) to process the loop iterations in parallel.

Until now, we have constructed a microarchitecture with a coarse-grained pipeline and a PE array that covers \textit{feature \#1} and part of \textit{feature \#2} of the CPP microarchitecture. Subsequently, we focus on loop scheduling inside PEs.

\noindent\textbf{4. Small loop flatten: } Based on our experiences, it is usually better to flat the in-PE loops with fixed, small trip counts. The reason is that 1) flatting loops with small trip counts provides more opportunities for HLS to generate a more efficient scheduling, and 2) flatting such loops will not affect the overall resource utilization considerably. As a result, we make an ad hoc strategy to fully unroll in-PE loops with trip count less than 16.

\begin{figure}
\begin{minipage}{\linewidth}
\begin{lstlisting}[basicstyle={\scriptsize\ttfamily},caption=NW Code with the CPP Microarchitecture, numbers=left,label=list:final]
void NW(...) {
  int M[129][129];
  // the array will be automatically partitioned
  ...
loop1: for(i=0; i<129; i++) { 
  #pragma Accel parallel factor=auto(1,128,seq)
    M[0][i] = ..,;     
  }
loop2: for(j=0; j<129; j++) { 
  #pragma Accel parallel factor=auto(1,128,seq)
    M[j][0] = ...; 
  }
loop3: for(i=1; i<129; i++) {
    for(j=1; j<129; j++) {...
    #pragma Accel parallel factor=auto(1,128,seq)
      M[i][j] = ...
  }}
  ...
}
void compute(char seqAs[], char seqBs[], char alignedAs[], char alignedBs[]) {
  for (int i=0; i<TILE_PAIRS; i++) {
  #pragma Accel parallel factor=auto(1,NUM_PAIRS,seq)
    NW(seqAs+i*128, seqBs+i*128, alignedAs+i*256, alignedBs+i*256);
}}
void load(...) { ... } // off-chip data load
void store(...) { ... } // off-chip data store
void kernel(char seqAs[], char seqBs[], char alignedAs[], char alignedBs[]) {
#pragma Accel bitwidth variable=seqAs factor=auto(8,512,pow2)
#pragma Accel bitwidth variable=seqBs factor=auto(8,512,pow2)
#pragma Accel bitwidth variable=alignedAs factor=auto(8,512,pow2)
#pragma Accel bitwidth variable=alignedBs factor=auto(8,512,pow2)

  char seqAs_buf_x[128*TILE_PAIRS]; char seqAs_buf_y[128*TILE_PAIRS];
  // the arrays will be automatically partitioned
  // the width will be automatically adjusted
  // the declarations for the other three buffers are omitted
  ...
  #pragma AutoAccel variable=TILE_PAIRS value=auto(1,NUM_PAIRS,seq)
  const int TILE_PAIRS = 16;
  int num_tiles = NUM_PAIRS/TILE_PAIRS;
  for (int i=0; i<num_tiles+2; i++) {
    if (i % 2 == 0) { 
      load(/* seqAs_buf_x <= seqAs, seqBs_buf_x <= seqBs */); 
      compute(seqAs_buf_y, seqBs_buf_y, alignedAs_buf_y, alignedBs_buf_y)
      store(/* alignedAs_buf_x <= alignedAs, alignedBs_buf_x <= alignedBs */); 
    }
    else { 
      load(/* seqAs_buf_y <= seqAs, seqBs_buf_y <= seqBs */); 
      compute(seqAs_buf_x, seqBs_buf_x, alignedAs_buf_x, alignedBs_buf_x)
      store(/* alignedAs_buf_y <= alignedAs, alignedBs_buf_y <= alignedBs */); 
}}}
\end{lstlisting} 
\end{minipage}
\end{figure}

\noindent\textbf{5. Fine-grained parallel/pipeline: } If an in-PE loop cannot be fully unrolled by step 4, it must satisfy one of the following conditions: 1) its trip count is either unknown or larger than 16, 2) it has loop carried-dependency, or 3) it contains one or more sub-loops that cannot be fully unrolled. In the first condition, we apply fine-grained parallelism and explore the best partial-unroll factor (lines 6, 10, and 15). In the other two conditions, we apply a fine-grained pipeline to improve the throughput and resource efficiency.

The above two transformations cover the remaining part of \textit{feature \#2}. Finally, we apply step 6 to cover \textit{feature \#3}.

\noindent\textbf{6. On-chip buffer reorganization: } We finally apply memory coalescing to reorganize the on-chip buffer (lines 28-31). We analyze the data type to determine the minimal bit-width, and always set the maximal bit-width to 512 bits since this is the maximal supported by the experimental platform. In addition, we only set the power-of-two bit-width values as DSE candidates, because HLS tools round BRAM sizes up to a power of two. As a result, this reduced design space can still cover the optimal solution in the original design space.

By applying the above code transformations, we are able to generate a transformed kernel code with the CPP microarchitecture and a design space.
As can be seen in Code~\ref{list:final}, the design space of the NW example has roughly $1.4\times10^{17}$ design points.
Therefore, an efficient DSE component is essential for the {\framework} framework.

\subsection{Design Space Exploration} \label{subsec:dse}
\begin{figure}[thb]
	\centering
	\includegraphics[scale=0.4]{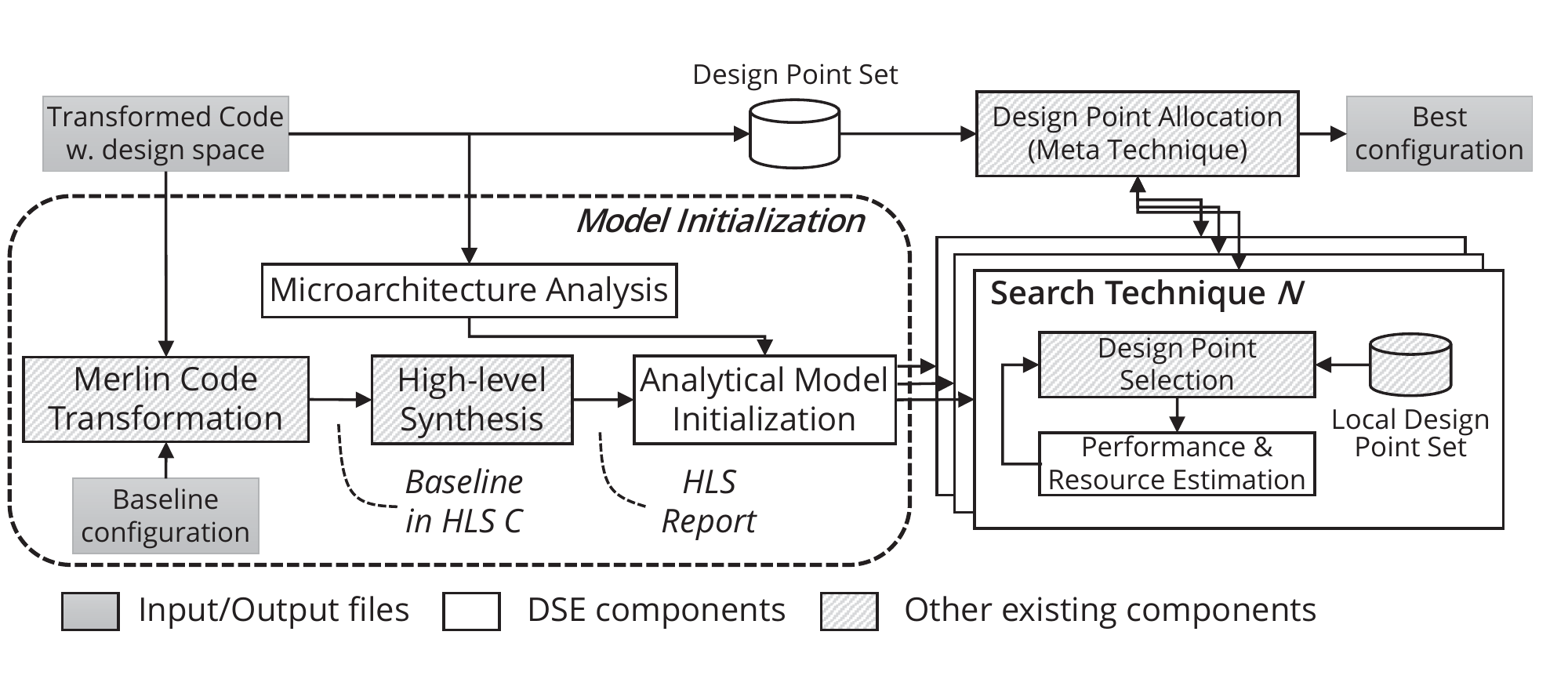} 
	\caption{Design Space Exploration Flow}
	\label{fig:model_dse}
\end{figure}

The DSE flow of {\framework}, as shown in Fig.~\ref{fig:model_dse}, is implemented using OpenTuner~\cite{opentuner}, an open-source framework for building domain-specific program tuners. The OpenTuner runtime has a search technique library that contains a collection of machine learning algorithms to cover as many customized tuning problems as possible. In order to assemble all search techniques, OpenTuner adopts a multi-armed bandit algorithm~\cite{Fialho2010} as a meta technique to judge the effectiveness of each search technique and allocate design points according to the judgment. Specifically, the search technique that can efficiently find high-quality design points will be rewarded and allocated more design points. In contrast, the technique that performs poorly on high-quality design point discovery will be allocated fewer points and eventually disabled. By harnessing OpenTuner, our DSE flow is able to realize the best design point efficiently and effectively.

In Figure~\ref{fig:model_dse}, the model initialization stage first parses the HLS reports of a few design points and generates the values of the design constants.
While most values are obtained by running HLS once, the LUT consumption of the standalone logic of a loop iteration ($R^{lut}_{r}$ in Eq.~\ref{eq:lut}) is calculated via two HLS reports.
In detail, we run HLS twice with two consecutive unroll factors of a loop to calculate the increment of the LUT consumption. This increment is the LUT consumption of the loop's standalone logic.
Next, we analyze the kernel source code to 1) establish the architecture hierarchy, and 2) fetch the design parameters and their value ranges from the \texttt{auto} pragmas inserted during design space establishment (Section~\ref{subsec:sda_create}). 
After the model is initialized, we simply feed the parameter sets of the remaining design points to the model and collect the performance and resource estimations.

%% file: 6_result.tex
\section{Experimental Evaluation} \label{sec:exp}
In this section we first describe our experimental setup, including hardware platform, software environments, and benchmarks. Then we evaluate {\framework} by analyzing the results of design space exploration (DSE), analytical model, and overall performance and energy efficiency.

\subsection{Experimental Setup} \label{subsec:exp_setup}
The evaluation of {\framework} is performed on the mainstream PCIe-based CPU-FPGA platform and the Xilinx SDAccel design flow. Table~\ref{tbl:exp_setup} lists the detailed hardware and software configuration. An Xeon CPU is connected with a Xilinx Virtex-7 FPGA board through the PCIe interface. For a fair comparison, both the CPU and the FPGA fabric were launched in 2012. On top of the platform hardware, we use Xilinx SDAccel to provide a hardware-software co-design environment.

Table~\ref{tbl:bench} lists the benchmarks used in our experiment. We use MachSuite~\cite{machsuite}, a benchmark suite that contains a broad class of computational kernels programmed as C functions for accelerator study, to evaluate the {\framework} framework. 
For each kernel, MachSuite provides at least one implementation that is based on a commonly used algorithm in software programming, which makes it a natural fit for demonstrating {\framework}.

\input{tables/exp_setup}
\input{tables/bench}



\subsection{Design Space Exploration Evaluation}
\label{subsec:dse_evaluate}
\begin{figure}[thb]
	\centering
	\includegraphics[scale=0.4]{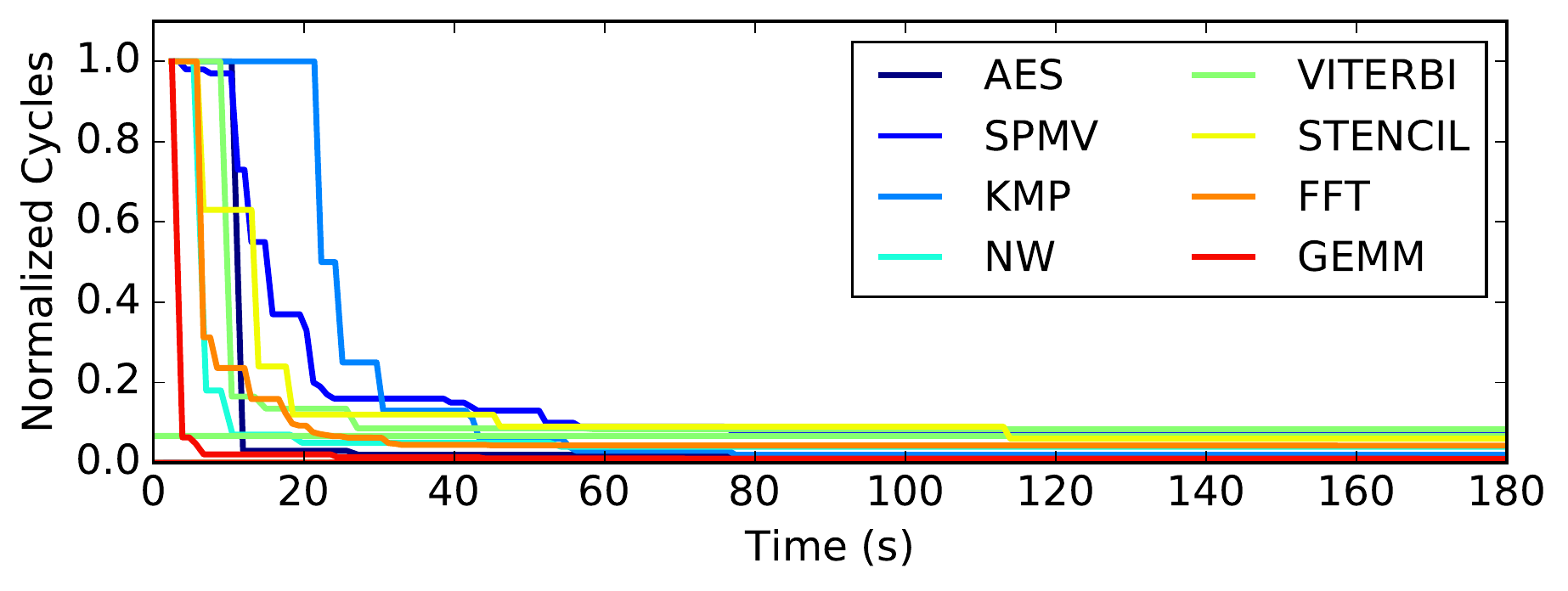} 
	\caption{Process of Finding the Optimal Design via DSE}
	\label{fig:exp_dse}
\end{figure}

Fig.~\ref{fig:exp_dse} illustrates the process of finding the optimal design point using the learning-based DSE approach with the analytical model to evaluate the performance and resource consumption. Thanks to the multi-armed bandit algorithm, the DSE process is able to find the right direction to the optimal solution efficiently, so the DSE time limit is set to only 180 seconds after the model initialization. As can be seen in Fig.~\ref{fig:exp_dse}, the execution cycles drop significantly in the first 20 seconds except for \texttt{KMP}. We analyze the process log of \texttt{KMP} in detail and find that the DSE spends some iterations  attempting to improve the performance of the compute stage, because \texttt{KMP} has a relative large design space inside the compute module. However, the performance of \texttt{KMP} is heavily bounded by memory bandwidth so reduced compute latency does not benefit for overall performance improvement. Despite this, the DSE process for \texttt{KMP} is still able to be converged in time. 

Based on the best configuration realized by the DSE, Table~\ref{tbl:result} presents the performance and resource utilization for each benchmark.\footnote{We set 80\% as the resource constraint based on the resources available for users.} Note that the \textit{C2C} metric in the second column is calculated by the following equation:

\begin{sequation}
C2C = \frac{Computation\ Cycle}{Communication\ Cycle}
\end{sequation}

\noindent The concept of C2C shares the merits of the CTC ratio used in~\cite{Zhang-fpga15}. 
We use C2C to analyze whether a design has achieved optimality ($C2C\sim1$). The design is identified as computational bound if C2C is larger than 1; otherwise it is communication bound. 

\input{tables/result}

According to Table~\ref{tbl:result}, the overall performance of \texttt{AES}, \texttt{SPMV}, \texttt{KMP}, and \texttt{STENCIL} is bounded by the off-chip bandwidth, because those four designs need to input or output a large amount of data, so the memory transaction time cannot be hidden by the computation time even if the {\framework} DSE has successfully found the design point with the largest bit-width. In fact, the memory-bounded design may potentially be further optimized by introducing data reuse analysis. For example, \cite{cong-tcad16} leverages polyhedral analysis to realize and optimize the data access pattern,  and this results in a much lower external memory transaction volume for stencil computation. However, the impact of this kind of transformation cannot be estimated by our analytical model and is beyond the scope of this paper. Future work would extend the model to cover those transformations.

For the other four designs, \texttt{VITERBI} and \texttt{NW} are bounded by LUTs. We can see that the C2C of both designs is higher than 2. It means that the PE in both designs consumes many LUTs, so even the overall design can still be further optimized by duplicating more PEs. There are no more available LUTs to use. 

On the other hand, \texttt{FFT} and \texttt{GEMM} are bounded by BRAM. Since their PE logics are relatively simple, BRAM becomes the major resource bottleneck. In this case, the DSE balances the computation and communication cycles by adjusting the PE number and buffer bit-width. As a result, the BRAM-bounded design has a C2C value larger than but close to 1.

\subsection{Analytical Model Evaluation}
We conduct two experiments to evaluate the accuracy of the analytical model.
The first experiment aims to evaluate whether the model-generated results are consistent with those collected from HLS reports.
In detail, we randomly select 20 design points for each benchmark, and compare the performance and resource usage for each design point between the model estimation and HLS report.
Table \ref{tbl:model} presents the average absolute difference rates for all cases.

We can see that the proposed model aligns to the HLS report accurately on performance and BRAM/DSP usage, and also results in only moderate differences on LUT/FF usage.
The differences are lead by the fact that the HLS tool adopts some resource-efficient implementations for its building blocks when a design requires a large proportion of on-board resources.
For example, \texttt{VITERBI} includes a loop statement with initiation interval (II) equal to 40.
The hardware circuit for this loop has some 25-to-1 multiplexers to select one floating-point number from 25 numbers.
We observe that when the number of PEs in the \texttt{VITERBI} design grows, the HLS tool automatically replaces a fully pipelined multiplexer implementation that consumes over 500 LUTs with the implementation that consumes only 32 LUTs to 1) meet the II=40 restriction, and 2) save on-board resources.
Since such dynamic optimization strategies are hard to capture with a static analytical model, a few percentages of differences on LUT/FF usage is inevitable. 

\input{tables/model_verify}


The second experiment evaluates the performance difference between the HLS report and the actual on-board result.
Table \ref{tbl:perf_err} presents the absolute performance difference rate of the optimal design point identified by {\framework}.
We can see that the average difference among all the benchmarks is only 6.2\%, which proves that the cycle estimation from the HLS tool is able to match the actual on-board execution time for the proposed microarchitecture. Note that the actual frequency of generated designs is not variant dramatically due to the following two reasons. First, Xilinx SDAccel 2016.3 optimizes the timing prior to optimizing other factors, so it might sacrifice resource efficiency (e.g., enlarge II) to preserve the frequency. Second, all of our designs reserve sufficient resources for the tool to avoid strict timing constraints. As a result, the impact of frequency on performance difference is moderate.

\input{tables/perf_err}

In addition, we further analyze the benchmarks with over 10\% performance difference, i.e., \texttt{AES} and \texttt{KMP}. We find that such relatively a large difference is mainly because the accelerator designs for these benchmarks have a very small execution time ($\sim$10 ms).
For these time frames, the start-up and end overhead bias the time significantly.
On the contrary, we also observe that the error rate of the model to on-board execution is always less than 5\% when a design has an over 100-millisecond execution time.
Hence, the proposed model is able to accurately predict the on-board execution time of a design given that its execution time is tens of milliseconds or larger.

\subsection{Performance and Energy Evaluation}
\begin{figure}[thb]
	\centering
	\includegraphics[scale=0.4]{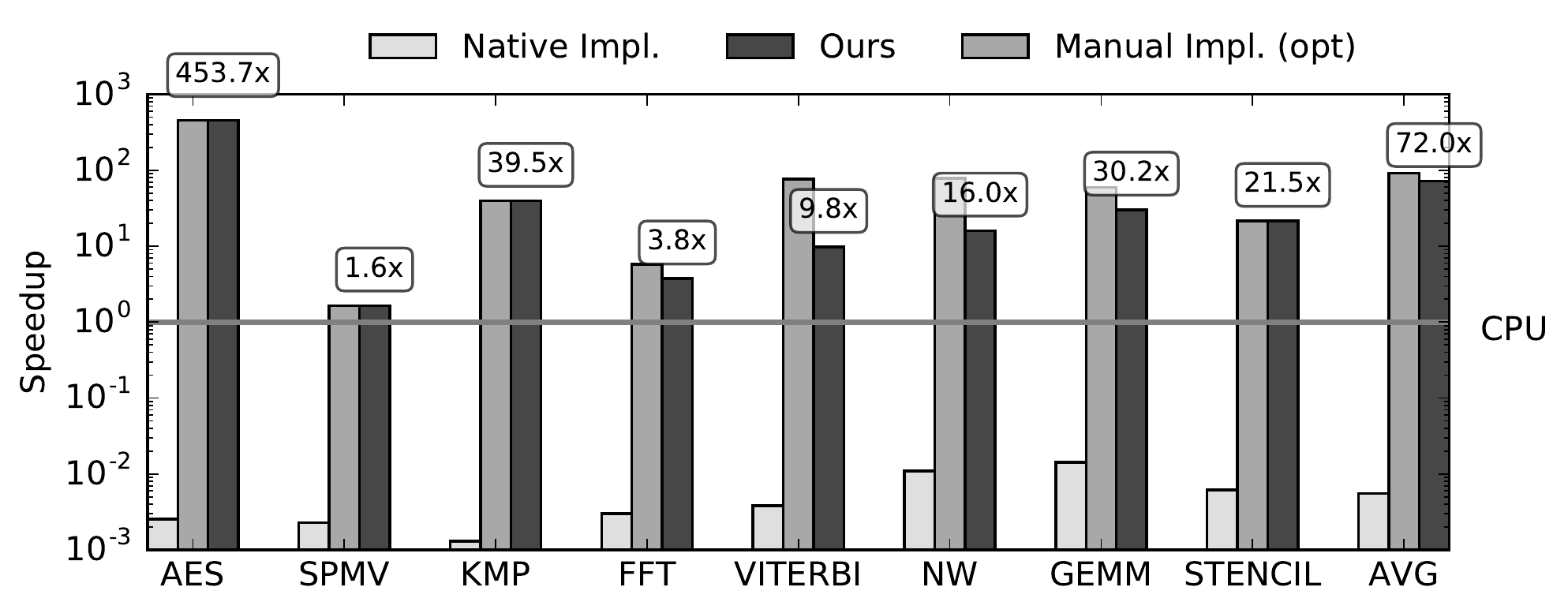} 
	\caption{Speedup over an Intel Xeon CPU Core}
	\label{fig:exp_cpu}
\end{figure}

We finally evaluate the performance speed-up and energy efficiency improvement of the generated FPGA accelerator designs. 
Figure~\ref{fig:exp_cpu} compares the performances between the naive implementation of MachSuite, manual HLS designs and {\framework}-generated accelerator designs, all of which are normalized to the performances of the corresponding software implementations.
We can clearly see that {\framework}-generated accelerators drastically outperform the naive implementations by {\avgimprove}, indicating that {\framework} has strongly addressed the gap from software programs towards high-quality hardware behavioral descriptions.
Meanwhile, the {\framework}-generated accelerators also outperform the software implementations by {\avgspeedup}, indicating that our approach does lead to high-quality accelerator designs.

We can also see from the experimental results that the manual designs only outperform the {\framework}-generated designs by an average 2.5$\times$, even after we spent several days to weeks applying more behavioral-level transformations to achieve the optimal performance.
In detail, for the benchmarks with \textit{C2C}\textless 1 --- \texttt{AES}, \texttt{SPMV}, \texttt{KMP} and \texttt{STENCIL} --- the generated designs have achieved the same optimal performance as manual designs in the experimental platform, because these benchmarks are all of linear time complexity, and their PEs run faster than the off-chip communication. On the other hand, the performances of the benchmarks with super-linear time complexity --- \texttt{FFT}, \texttt{NW}, \texttt{VITERBI} and \texttt{GEMM} --- are bounded by FPGA on-chip resources. As a result. the performance can potentially be further improved by using application-specific accelerator circuits to improve resource efficiency.
For example, we use the systolic array microarchitecture to improve the GEMM accelerator design and achieve the optimal performance with all on-chip DSPs.
Although such specialized architectures cannot be covered by {\framework}, {\framework} still preserves high accelerator quality while substantially improving the FPGA programmability. 

Finally, we analyze the energy efficiency gain of {\framework}-generated designs. We estimate the energy efficiency (performance per watt) of our experiments by considering execution time and thermal design power (TDP). The TDP of the Intel Xeon CPU and the Xilinx FPGA used in this comparison is 80W and 25W, respectively. Accordingly, {\framework}-generated designs can achieve up to 1677.9$\times$ energy efficiency improvement and 260.4$\times$ on average. 

%% file: tables/exp_setup.tex
\begin{table}[thb]
\centering
\scriptsize
\caption{Configuration of Hardware and Software}
\label{tbl:exp_setup}

\begin{tabular}{|l|l|}
\hline
Host CPU Model & Intel Xeon E5-2420 @ 1.9GHz (released in 2012) \\ \hline
Host Memory & 64GB DDR3-1600 \\ \hline \hline
FPGA Fabric & Xilinx Virtex-7 @ 200MHz (released in 2012) \\ \hline
Device Memory & 8GB DDR3-1600 (Max Band.: 12.8GB/s) \\ \hline \hline
CPU-FPGA Interface & PCIe Gen3 x8 (Max Band.: 8GB/s) \\ \hline
Transformation Flow & Merlin compiler 2017.1 \\ \hline
Synthesis Flow & SDAccel 2016.3 \\ \hline
\end{tabular}
\end{table}

%% file: tables/bench.tex
\begin{table}[thb]
  \centering
  \caption{Benchmark Description}
  {\scriptsize
  \begin{tabular}{|l|l|}
\hline
Kernel  & Description and Input Information     
\\ \hline \hline
AES     & Advanced encryption standard. Input: 256-bit key; 64MB data.                                                                              \\ \hline
GEMM    & \begin{tabular}[c]{@{}l@{}}General matrix multiplication ($O(N^{3})$).\\ Input: two 1024$\times$1024 64-bit floating-point matrices\end{tabular} \\ \hline
KMP     & \begin{tabular}[c]{@{}l@{}}Knuth-Morris-Pratt string matching.\\ Input: 128MB string; 16B substring.\end{tabular}                         \\ \hline
NW      & \begin{tabular}[c]{@{}l@{}}Needleman-Wunsch sequence alignment.\\ Input: 64K pairs of 128-nucleotide seq.\end{tabular}                    \\ \hline
SPMV    & \begin{tabular}[c]{@{}l@{}}Sparse matrix-vector multiplication.\\ Input: 4096$\times$512 ELLPACK data and index.\end{tabular}             \\ \hline
VITERBI & Viterbi algorithm. Input: 1M 128-element chains.                                                                                          \\ \hline
FFT     & Fast Fourier transform. Input: 65536 strides each with 1KB size.                                                                          \\ \hline
STENCIL & Stencil computation. Input: a 4096$\times$4096 image                                                                                             \\ \hline
\end{tabular}
  }
\label{tbl:bench}
\end{table}


%% file: tables/result.tex
\begin{table}[thb]
\centering
\caption{C2C and Resource Utilization}
\label{tbl:result}
{\scriptsize
\begin{tabular}{|l|ccccc|}
\hline
Bench.  & C2C                   & BRAM                           & LUT                            & DSP    & FF     \\ \hline\hline
AES     & {\ul \textit{\textbf{0.4}}} & 27.3\%                         & 23.1\%                         & 0.0\%  & 3.2\%  \\ \hline
SPMV    & {\ul \textit{\textbf{0.6}}} & 43.0\%                         & 5.8\%                          & 4.3\%  & 3.2\%  \\ \hline
KMP     & {\ul \textit{\textbf{0.1}}} & 52.4\%                         & 14.0\%                         & 0.0\%  & 1.8\%  \\ \hline
FFT     & 1.1                         & {\ul \textit{\textbf{78.5\%}}} & 48.8\%                         & 66.7\% & 24.2\% \\ \hline
VITERBI & 2.2                         & 20.4\%                         & {\ul \textit{\textbf{79.4\%}}} & 12.6\% & 21.8\% \\ \hline
NW      & 4.0                         & 17.2\%                         & {\ul \textit{\textbf{78.4\%}}} & 0.0\%  & 25.9\% \\ \hline
STENCIL & {\ul \textit{\textbf{0.5}}} & 52.3\%                         & 5.1\%                          & 24.8\% & 2.3\%  \\ \hline
GEMM    & 1.1                         & {\ul \textit{\textbf{74.4\%}}} & 32.5\%                         & 49.8\% & 21.2\% \\ \hline
\end{tabular}
}
\end{table}

%% file: tables/model_verify.tex
\begin{table}[thb]
\centering
\caption{Differences Between Model and HLS Reports}
\label{tbl:model}
{\scriptsize
\begin{tabular}{|c|ccccc|}
\hline
Parameter 	& Perf. & BRAM & DSP & LUT & FF \\ \hline\hline
Avg. error 	& \textless 1\%  & \textless 1\%     & \textless 1\%  	 &  6.5\%   & 4.3\%    \\ \hline
\end{tabular}
}
\end{table}

%% file: tables/perf_err.tex
\begin{table}[thb]
\centering
\caption{Differences Between Model and On-board Results}
\label{tbl:perf_err}
{\scriptsize
\begin{tabular}{|c|cccc|}
\hline
Bench.    & AES     & SPMV  & KMP     & FFT   \\ \hline
\hline 
Avg. err. & 13.5\%  & 9.5\% & 12.2\%  & 0.1\% \\ \hline
\hline
Bench.    & VITERBI & NW    & STENCIL & GEMM  \\ \hline
\hline
Avg. err. & 2.1\%   & 1.1\% & 7.7\%   & 3.3\% \\ \hline
\end{tabular}
}
\end{table}

%% file: 7_related.tex
\section{Related Work} \label{sec:related}
In this section we discuss related work in the analytical models and the automated frameworks for FPGA design optimization.


\noindent\textbf{Analytical Modeling:} Fast performance estimation on FPGAs has become popular in recent years. In general, performance analysis is mainly performed at either IR level~\cite{fpga-opencl,zhongdesign,raghu-asplos,dhdl,gao-fpga16} or source code level~\cite{zohouri16}. 
Since most of the existing work performs analysis without explicitly considering back-end design flow~\cite{fpga-opencl,zhongdesign,gao-fpga16,raghu-asplos,dhdl}, their analysis cannot reflect the optimization done by the commercial tool. On the other hand, similar to this paper, \cite{zohouri16} builds the performance model with the help of the commercial tool, but \cite{zohouri16} provides neither the resource model nor automated code transformation, so users still need to manually change the kernel code while considering the FPGA resource limitation.


\noindent\textbf{Automated Framework:} Some projects aim to provide an automated framework to perform code generation and design space exploration~\cite{raghu-asplos,dhdl,melia}. The framework presented by \cite{raghu-asplos,dhdl} accepts parallel patterns (e.g., map, groupBy, filter, reduce, etc.) and performs FPGA accelerator generation with analytical DSE.
Different from this paper, which automatically applies the CPP microarchitecture, the FPGA architecture generated by \cite{raghu-asplos,dhdl} is composed of predefined hardware components (i.e., memory, controller, and primitive operations) to guarantee efficiency. However the selection of these components highly depends on the semantic information of user-specified parallel patterns. Furthermore, the performance model for DSE in \cite{raghu-asplos,dhdl} is built only for the predefined hardware components.

In addition, Melia~\cite{melia} is a MapReduce framework that supports automated code generation from user-written C code to OpenCL. Melia asks users to provide the best configuration by leveraging the model from \cite{fpga-opencl} to generate the OpenCL code. Consequently, Melia only generates the FPGA accelerator design under a MapReduce programming model, and misses automatic design space exploration.

Finally, some frameworks also focus on general-purpose programming languages such as C/C++~\cite{li-fpga15,lin-analyzer,gao-fpga16}. SOAP3~\cite{gao-fpga16} is a framework that analyzes a kernel at the metasemantic intermediate representation (MIR) graph level and transforms it according to the result of design space exploration. However, SOAP3 adopts regression models for resource estimation, so the model is not general enough to cover nonlinear resource consumption. A framework in \cite{li-fpga15} uses an analytical model based on HLS results (like this paper) for maximizing throughput given resource constraints. However, they consider only loop pipelining and ignore the design space of coarse- and fine-grained parallelism. Lin-analyzer~\cite{lin-analyzer} is a framework to identify the performance bottleneck for C/C++ programs, but it does not involve code transformation and only focuses on fine-grained parallelism.


%% file: 8_conclusion.tex
\section{Conclusion} \label{sec:conclusion}
While the FPGA-based heterogeneous architectures are becoming a promising paradigm to provide continued performance and energy improvement in modern datacenters, accelerator programming arises as a serious challenge to application developers.
In this paper we propose the {\framework} framework to provide a nearly push-button experience on mapping C functions into high-quality FPGA accelerator designs.
Featuring the CPP microarchitecture, a fast analytical model-based design space exploration and automatic code transformation, {\framework} achieves {\avgspeedup} speed-up and 260.4$\times$ energy improvement for a broad class of computation kernels.

Furthermore, we believe that the design principles of {\framework} can be further generalized to stimulate more research on the adoption of FPGAs in datacenters. 
For example, the CPP microarchitecture serves as a proof-of-concept that using an accelerator design template as a specification of the program-to-behavioral-description transformation drastically reduces the design space while preserving the accelerator quality.
Therefore, more microarchitectures might be added in {\framework} to improve the coverage of computation kernels.
Also, more sophisticated, high-abstract code transformations  (e.g., loop permutation) are able to be supported in the future, along with polyhedral analysis, to form a larger design space and create more optimization opportunities.